**Lateral migration of living cells in inertial microfluidic systems explored by fully three-dimensional numerical simulation**

Hongzhi Lan [a], Soojung Claire Hur [b], Dino Di Carlo [c], and Damir B. Khismatullin [a,*]

[a] *Department of Biomedical Engineering, Tulane University, New Orleans, LA, USA*

[b] *Rowland Institute at Harvard University, Cambridge, MA, USA*

[c] *Department of Biomedical Engineering, University of California, Los Angeles, CA, USA*

Running head: Lateral migration of cells in a microchannel

___________________

[*]Address for correspondence: Dr. Damir B. Khismatullin, Department of Biomedical Engineering, Tulane University, New Orleans, LA 70118, USA. Tel.: +1 504 247 1587; Fax: +1 504 862 8779; E-mail: damir@tulane.edu.

**Abstract.** The effects of cell size and deformability on the lateral migration and deformation of living cells flowing through a rectangular microchannel has been numerically investigated and compared with the inertial-microfluidics data on detection and separation of cells. The results of this work indicate that the cells move closer to the centerline if they are bigger and/or more deformable and that their equilibrium position is largely determined by the solvent (cytosol) viscosity, which is much less than the polymer (cytoskeleton) viscosity measured in most rheological systems. Simulations also suggest that decreasing channel dimensions leads to larger differences in equilibrium position for particles of different viscoelastic properties, giving design guidance for the next generation of microfluidic cell separation chips.

Inertial microfluidics has recently emerged as a new tool for the detection and separation of cells with different sizes and mechanical properties [1-5]. In this approach, fluid dynamic effects during perfusion through a microchannel lead to drift of living cells and other deformable particles to specific lateral equilibrium positions depending on their size and deformability [6-9]. Traditionally, cell sorting in microfluidic systems requires the application of an external electromagnetic force field or fluorescent markers [10]. The use of purely hydrodynamic and label-free particle manipulation methods would enable a high-throughput and low-cost way for cell sorting and disease diagnosis for a number of pathological conditions (sepsis, cancer metastasis, etc.) associated with changes in cell deformability [5,11,12].

Here, we explore numerically the effect of size and bulk viscoelasticity on lateral migration and deformation of small deformable particles in a rectangular duct. This problem has not been considered in previous numerical work. The emphasis in the literature to date has been placed on low-Reynolds-number flows, large liquid drops or elastic capsules, and parallel-plate or tube geometries [8,13-22]. The results of our numerical study are compared with the experimental data obtained using a microchannel with identical dimensions [1].

Our numerical model is based on a fully three-dimensional algorithm for two-phase viscoelastic flows, where the fluid-fluid interface is tracked by the volume-of-fluid (VOF) method [22,23]. The disperse phase (i.e., drop or cell) is a complex fluid characterized by its own internal compartments: a solvent (cytosol) and a polymer matrix (cytoskeleton). The continuous phase (external fluid) is a Newtonian fluid. We assume that the solvent viscosity $\mu_s$ is equal to the viscosity of the continuous phase, which is water in this simulation. The velocity field in different phases is solved by one set of the Navier-Stokes equations with the values of physical parameters (mass density, shear viscosity, etc.) averaged over each grid cell with multiple phases [22]:

$$\nabla \cdot \mathbf{u} = 0, \tag{1}$$

$$\rho\left(\frac{\partial \mathbf{u}}{\partial t} + \mathbf{u} \cdot \nabla \mathbf{u}\right) = \nabla \cdot \mathbf{T} - \nabla p + \nabla \cdot (\mu_s (\nabla \mathbf{u} + (\nabla \mathbf{u})^T)) + \mathbf{F}. \tag{2}$$

Here, $\mathbf{u}$ is the velocity vector, $\rho$ the mass density, $\mathbf{T}$ the extra stress tensor that represents the polymer matrix (cytoskeleton) contribution to the shear stress field inside the disperse phase; and $p$ pressure. The net body force $\mathbf{F}$ per unit volume includes gravity $\rho \mathbf{g}$ and the interfacial tension force $\mathbf{F}_t$ calculated by the continuous surface force (CSF) method [24] as follows:

$$\mathbf{F}_t = \sigma \tilde{\kappa} \|\nabla c\| \hat{\mathbf{n}}, \tag{3}$$

where $\sigma$ is the interfacial tension coefficient, $\hat{\mathbf{n}} = \nabla c / \|\nabla c\|$ the outward unit normal, $\tilde{\kappa} = -\nabla \cdot \hat{\mathbf{n}}$ the mean curvature, $c = c(t, \mathbf{x})$ the concentration function (color function) that takes the value between 0 and 1 at the interface, and $\mathbf{x} = (x, y, z)$ the position vector in the Cartesian coordinate system. The Navier–Stokes equations were solved by the projection method on a staggered Marker-and-Cell grid with a predictor–corrector semi-implicit factorized scheme for the intermediate velocity approach. Details about this approach are given in our previous report [25]. We have not observed any significant changes in the lateral equilibrium position for particles with the density difference as large as 200 kg/m$^3$. The density of living cells such as leukocytes (white blood cells) is about 80 kg/m$^3$ greater than the density of water [26]. To simplify the problem, we assume that the mass density of the external fluid and the cell are equal.

The viscoelasticity of the disperse phase was described mathematically by the Giesekus constitutive equation [22]:

$$\lambda_1\left(\frac{\partial \mathbf{T}}{\partial t} + (\mathbf{u}\cdot\nabla)\mathbf{T} - (\nabla\mathbf{u})\mathbf{T} - \mathbf{T}(\nabla\mathbf{u})^T\right) + \mathbf{T} + \lambda_1 \kappa \mathbf{T}^2 = \lambda_1 G(0)(\nabla\mathbf{u} + (\nabla\mathbf{u})^T). \tag{4}$$

Here $\kappa$ the Giesekus nonlinear parameter; $G(0)$ the elastic modulus at $t = 0$; $\lambda_1$ the relaxation time, i.e., the ratio of the polymer viscosity $\mu_p$ to $G(0)$.

In the simulation, we consider a deformable particle of initially spherical shape with diameter $D$ suspended below the centerline in a rectangular microchannel with height $H = 38$ μm and width $W = 85$ μm (Fig. 1; width $W$ is the channel size in $y$ direction not shown in this figure). The particle was initially placed with its center on the vertical mid-plane $y = W/2$. The flow is fully established with the centerline velocity $U_{max} = 0.4$ m/s, according to the experimental setup of Hur et al. [1]. All the walls of the rectangular duct are considered to be no slip boundaries. During the simulation, the particle migrates laterally (in the $z$-direction) to some equilibrium position $z_{eq}$ independently of its initial location. Fluid shear stresses also induce the particle elongation along the principal axis of deformation with an acute angle $\theta$ between this axis and the $x$-axis (Fig. 1). We quantify the particle deformation by the Taylor deformation index [27]:

$$D_{xz} = \frac{D_{max} - D_{min}}{D_{max} + D_{min}}, \tag{5}$$

where $D_{max}$ and $D_{min}$ are the lengths of the drop in the $xz$-plane along the long and short principal axes of deformation.

To properly understand the lateral migration properties of living cells, we first need to analyze the migration of Newtonian liquid drops, where the extra stress tensor $\mathbf{T}$ is zero. Figure 2 shows both numerical and experimental data on the equilibrium position and deformation of the drops for different bulk viscosities and sizes. The details on drop migration experiments can be found elsewhere [1]. According to the numerical simulation, for a smaller drop with $D/H = 0.2$, the dimensionless lateral equilibrium position $2z_{eq}/H$ decreases monotonically from 0.56 to 0.47 as the drop viscosity increases from 0.83 cP to 50 cP (open circles in Fig. 2(a)). The $2z_{eq}/H$ equal to 0 and 1 indicates that the center of the drop is located at the channel wall and the centerline, respectively. This trend is consistent with the experimental data (solid squares), excluding a small region from 0.83 cP to 9.3 cP, where the equilibrium position increases slightly from 0.6 to 0.61. In the limit of non-deformable (rigid) particles, the equilibrium position reduces to 0.44 according to experiments, which is only 7% lower than the value of 0.47 predicted by the model for 50 cP-viscosity drops. It should be noted these drops experience very small deformation (cf. the third image in Fig 2(e)). Their deformation index is only 0.035 (Fig. 2(c)). One explanation for the discrepancy between the model and experiment is that drops in experiments might have a lower interfacial tension than the estimate used in experimental data analysis [1]. For example, our previous numerical analysis of lateral migration of small drops ($D/H = 0.2$) indicates that a decrease in interfacial tension from 0.5 mN/m to 0.1 mN/m reverses the dependence of the drop equilibrium position on the drop viscosity [25]. The effects of the Reynolds number and the capillary number on the equilibrium position of capsules in a micro-channel were also modeled in [28].

Figure 2(b) shows the effect of drop size on drop lateral migration. The numerical model (dashed lines) predicts an almost linear increase in the equilibrium position with the drop-to-channel height ratio increased from 0.2 and 0.5: from 0.52 to 0.78 at a viscosity of 9.3 cP (open circles) and from 0.47 to 0.62 at 50 cP (crosses). Thus, large drops with low bulk viscosity approach the position close to the centerline. This result also indicates that the size-based separation of particles in a microfluidic system becomes easier with increased deformability of particles; and vice versa, deformability-based separation becomes easier for larger particles.

There is a good agreement between simulation and experiment for drops with viscosity of 9.3 cP (compare open and solid circles in Fig. 2(b)). Deviation increases with a decrease in *D/H*. Drop deformation was not quantified in experiments and, therefore, Figs. 2(c-e) include only the numerical data for the deformation index, orientation angle, and shape of drops with *D/H*=0.2 and different viscosities. The deformation index decreases from 0.17 to 0.04 and the orientation angle decreases from 37º to 3º when the drop viscosity increases from 0.83 cP to 50 cP. This change in shape could be detected from the images of flowing particles in experiments, thus allowing the estimation of the corresponding change in the deformability of these particles. These results indicate that increased deformation and angle of the deformed shape of a droplet correlate with increased center-directed migration. The connection between drop shape and lift has been previously posed by Abkarian et al. [29] as well.

Many living cells can be modeled as viscoelastic drops with interfacial tension (known in this case as cortical tension) produced by the elasticity of the actin cytoskeleton in the cell cortex [30]. Such models were specifically used for the determination of rheological properties of human neutrophils and other leukocytes by micropipette aspiration [22,31-34]. In this study, we consider cortical tension and model the viscoelasticity of living cells by the Giesekua constitutive equation (Eq. (4)). Based on experimental measurements, the range of cytoplasmic viscosity is between 65 P and 2100 P, the cortical tension coefficient is 30 pN/µm, and the relaxation time is about 0.2 s [30,32,35-39]. These values (our polymer viscosity values were 100 P, 1000 P, and 10000 P) have been considered in the simulation.

It is important to emphasize that the apparent cytoplasmic viscosity measured in experiments is not equivalent to the solvent or polymer viscosity in our model. When $\kappa = 0$ (as in the current simulation) and the shear rate is constant, the apparent viscosity $\mu^+$ of the viscoelastic drop changes with time as follows:

$$\mu^+ = (\mu_s + \mu_p)\left[\frac{\mu_s}{\mu_s + \mu_p} + \left(1 - \frac{\mu_s}{\mu_s + \mu_p}\right)(1 - e^{-t/\lambda_1})\right] \tag{6}$$

It follows from Eq. (6) that the apparent viscosity is initially equal to the solvent viscosity $\mu_s$ and approaches the polymer viscosity $\mu_p$ when time $t$ becomes much larger than the relaxation time $\lambda_1$ (cf. Fig. 3).

Figure 4 shows both the numerical data and experimental results on the lateral migration of living cells with different viscoelasticity and size (breast cancer cell lines MCF7 and modMCF7 and white blood cells from human blood). MCF7 are benign breast cancer cells, which were found to be less deformable than chemically modified modMCF7 that have an increased metastatic potential [5,40]. See details of these experiments in Hur et al. [1].

According to the numerical simulation, if the relaxation time is fixed, the cell equilibrium position moves closer to the wall with increasing polymer viscosity (dashed lines in Fig. 4(a)). When the polymer viscosity changes from 100 P to 10000 P, the equilibrium position of the cell with D/H = 0.2 and 0.5 changes from 0.72 to 0.48 and from 1.0 to 0.67, respectively. As in the case of drops, larger cells move closer to the centerline. These trends are consistent with the experimental data (inverted triangles, solid circles, and solid squares in Fig. 4(a)). Less deformable MCF7 (solid circles) have an equilibrium position closer to the wall than modMCF7 (solid squares). From the comparison of the simulation and experiment, the polymer viscosity of the cells studied falls within the range chosen in the simulation (100-10000 P). The cells with polymer viscosity of 10000 P (asterisks) behave similarly to rigid particles (solid diamonds) or 50 cP-viscosity drops (crosses) in Fig. 2(b). On the other hand, the cells with polymer viscosity of 100 P (crosses) have an equilibrium position reaching the centerline at $D/H > 0.4$. From comparison of the numerical and experimental data in Fig. 4(a), it follows that the polymer viscosity of white blood cells and MCF7 cells is slightly higher than 1000 P, while it is between 500 P and 1000 P for modMCF7 cells. The majority of micropipette aspiration measurements also predict the values close to 1000 P for the cytoplasmic viscosity of white blood cells [34-36]. It should be noted that the cells exposed to conditions of in vitro experiments can be easily activated and thus become less deformable. Therefore, leukocytes may experience much higher deformation in vivo than in vitro [30].

Interestingly, the cells (and viscoelastic drops) with the polymer viscosity less than 10000 P have an equilibrium position closer to the centerline than Newtonian liquid drops with much lower viscosity (50 cP). This seemingly conflicting observation can be explained by Fig. 3. In experiments and modeling, the migration and deformation of cells happens within a very short time (a few milliseconds). According to Fig. 3, the apparent cytoplasmic viscosity is still very close to the solvent (water) viscosity during this period. This is especially true when the elastic modulus $G(0)$ is sufficiently small. In this situation, cortical tension still contributes to the lateral migration and since it is much less for cells than the interfacial tension for drops, the cells go to a higher equilibrium position than drops with similar internal viscosity. Additionally, the cells with polymer viscosity of 100 P in Fig. 4(b) experience larger deformation than liquid drops with viscosity of 0.83 cP in Fig. 2(c). This is again the result of the reduced apparent viscosity of cells during the migration time period.

Another interesting observation is a shift in the equilibrium position of cells closer to the wall from 0.72 to 0.61 at very small values of the relaxation time (0.02 s and 0.0002 s in Fig. 4(c), respectively). In this simulation, the elastic modulus $G(0)$ was kept fixed. No significant change in the apparent viscosity is observed between $\lambda_1 = 0.02$ s and 2.0 s, but the effect of relaxation time becomes significant at $\lambda_1 < 0.02$ s. Since a cell with a very low value of the relaxation time (0.0002 s) and fixed elastic modulus (50 Pa) behaves essentially as a low-viscosity (11 cP) liquid drop, its equilibrium position approaches the value of 0.61 that characterizes liquid drops with low viscosity and interfacial tension studied in experiments (Fig. 2(a). If we fix the polymer viscosity instead of the elastic modulus, a decrease in the relaxation time means an increase in the elastic modulus and, as a result, the cell becomes stiffer and its equilibrium position becomes closer to the wall. As seen in Fig. 4(d), an increase in $G(0)$ from 50 Pa to 50000 Pa moves the equilibrium position from 0.72 to 0.47.

The lateral migration and deformation of living cells and liquid drops in a rectangular microchannel has been studied numerically. The results show that the size and viscoelasticity of cells influences their lateral equilibrium position, which are consistent with recent experimental work by Hur et al. [1]. Cells move closer to the centerline if they are bigger and/or more deformable. This observation enables inertial microfluidics-based detection and separation of cells with different deformability and size. Our study also indicates that proper attention should be given to the effect of the apparent viscosity increase over time during lateral migration of cells when designing inertial microfluidics systems for the determination of rheological properties of cells. If the migration time period is much less than the relaxation time of the cell, the equilibrium position of the cell will be determined by the solvent viscosity, which is much less than the polymer viscosity or apparent cytoplasmic viscosity measured in other rheological systems. Our results also emphasize that channel geometry can be tuned to assist in deformability-based separations: smaller channel dimensions lead to larger differences in the equilibrium position for cells or drops of the same size but with different viscoelastic properties.

**Acknowledgements**

This work was supported by Louisiana Board of Regents grant LEQSF(2011-14)-RD-A-24 to D.K., the LONI Institute Graduate Fellowship and the Tulane-IBM Corporation Fellowship in Computational Science to H.L., and NSF Grant # 0930501 to D.D.C.

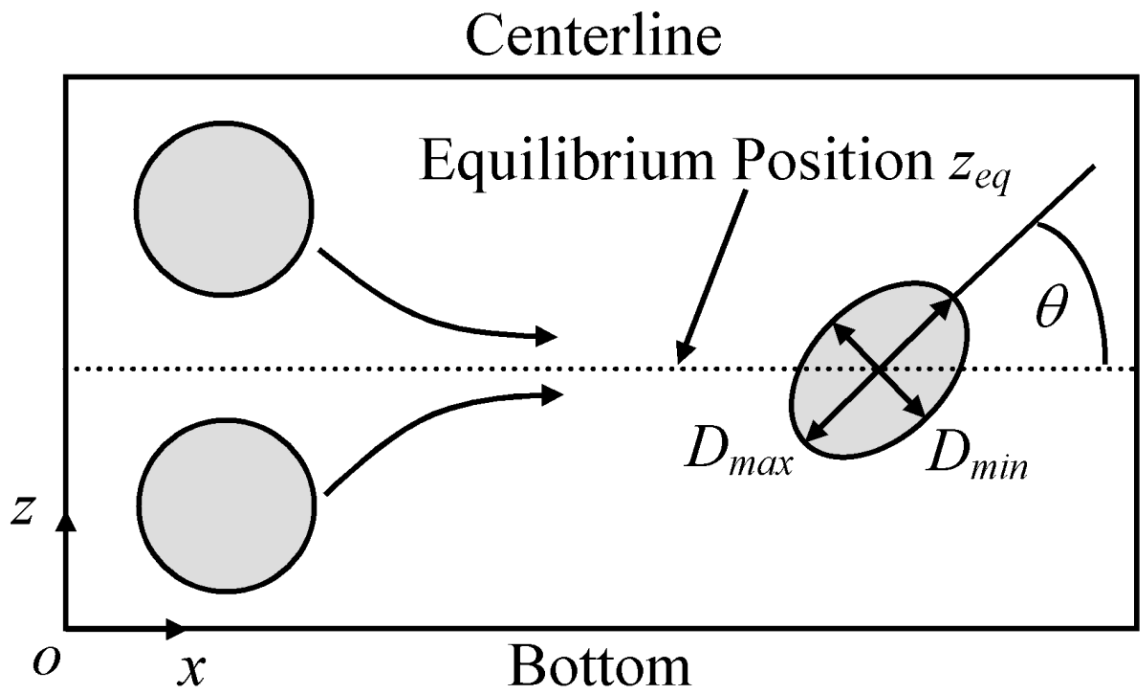

Fig. 1. Schematic of lateral migration and deformation of a deformable particle in a microchannel.

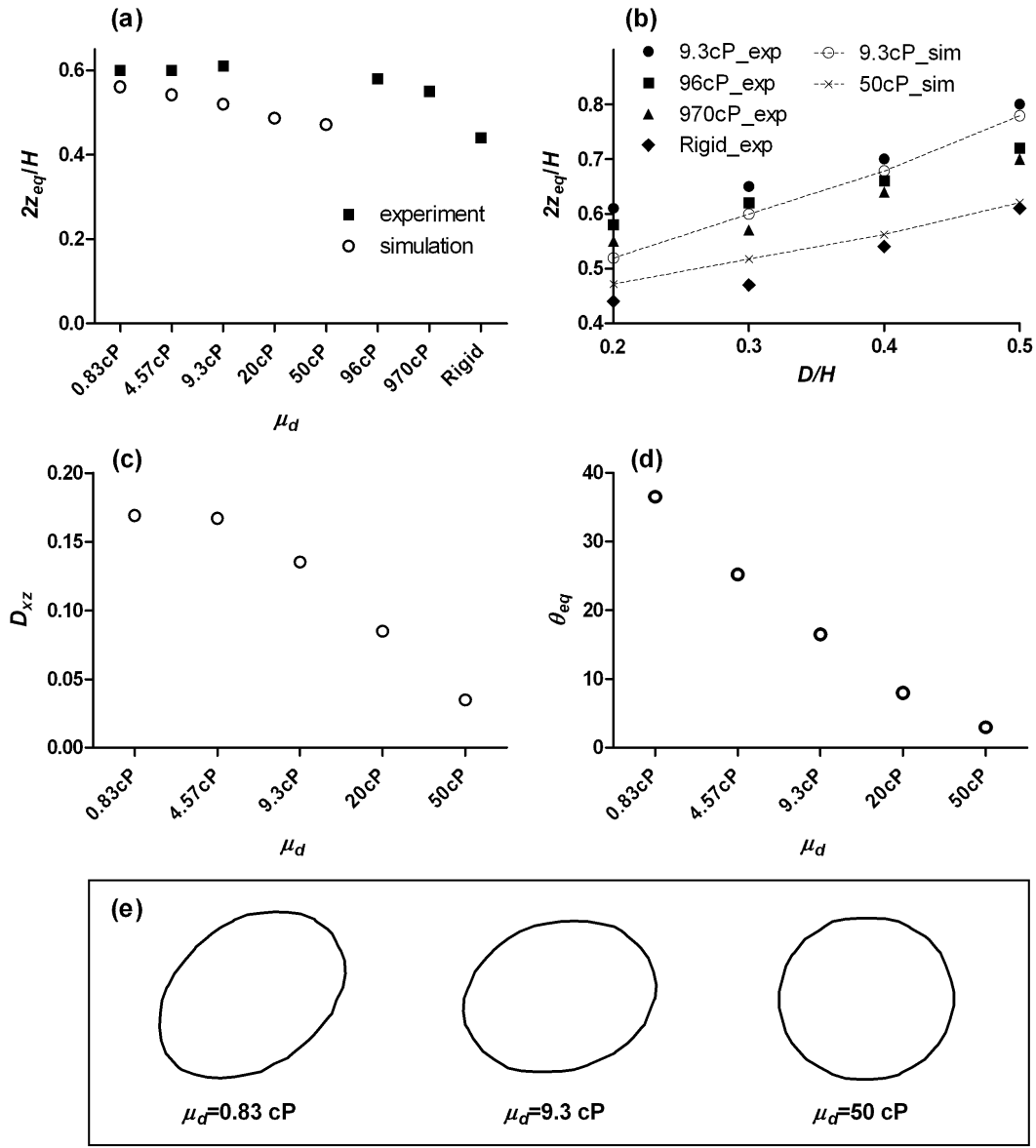

Fig. 2. (a) Comparison of equilibrium position of drops ($D/H$=0.2) with various viscosities $\mu_d$ between our simulation (open circles) and experimental results (mean values, solid squares). (b) Comparison of the equilibrium position of drops with different sizes between our simulation (dashed lines) and empirical results (solid symbols are the mean values). (c) Simulation results for the deformation of drops ($D/H$=0.2) with different viscosity. (d) Simulation results for the orientation angle of drops ($D/H$=0.2) with different viscosity. $\sigma$=0.5 mN/m. (e) Deformed shapes of drops ($D/H$=0.2) with different viscosity, according to numerical simulation.

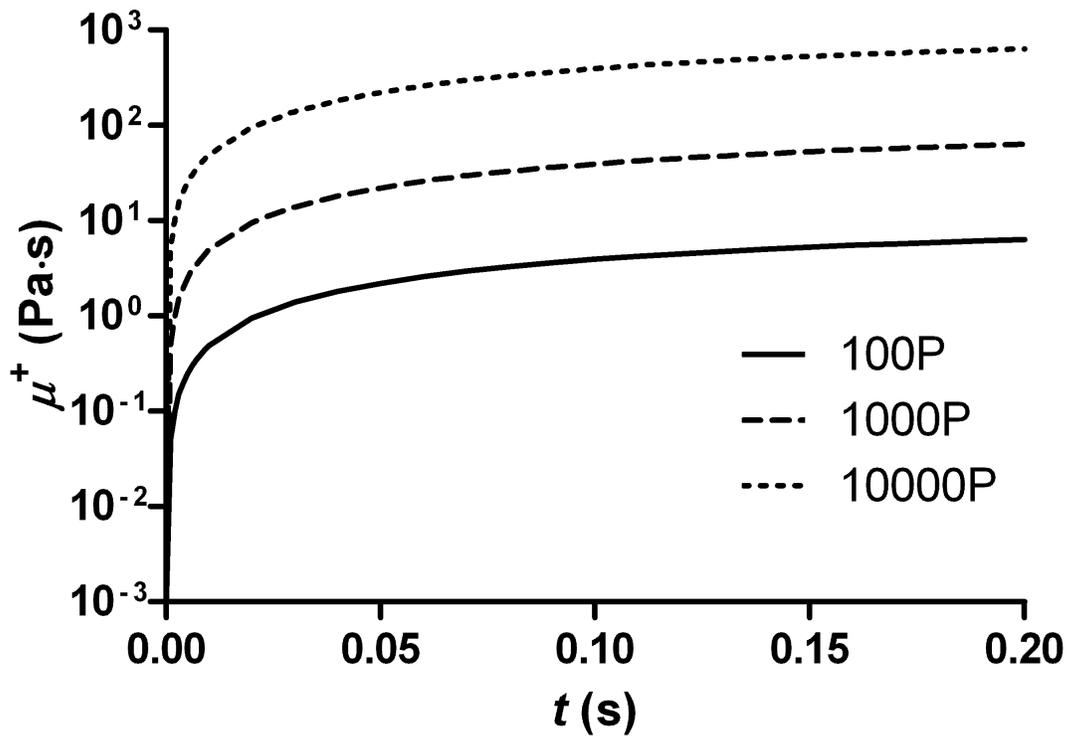

Fig. 3. Apparent viscosity with different initial elastic modulus $G(0)$ ( 50, 500, 5000 Pa) or polymer viscosity $\mu_p$ (100, 1000, 10000 P). Relaxation time $\lambda_1$ is fixed at 0.2 s.

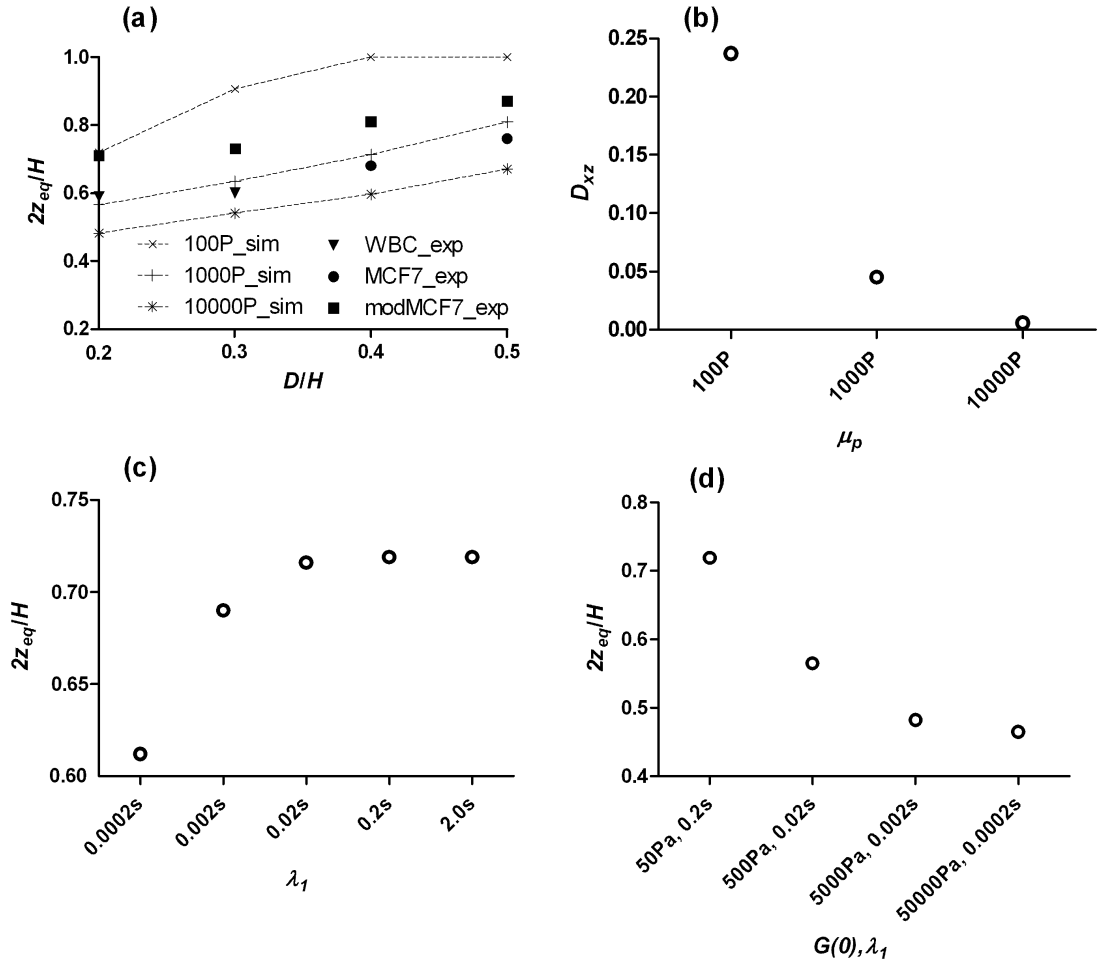

Fig. 4. (a) Comparison of equilibrium position of viscoelastic drop and cells with different sizes between our simulation (dashed lines) and the experimental results of Hur et al. (b) Simulation results for the deformation of viscoelastic drops with different polymer viscosity ($D/H$=0.2, $\lambda_1$=0.2 s). (c) Simulation results for the equilibrium position of viscoelastic drops with different relaxation time $\lambda_1$ and constant initial elastic modulus $G(0)$ ($D/H$=0.2, $G(0)$=50 Pa). (d) Simulation results for the equilibrium position of viscoelastic drops with different relaxation time $\lambda_1$ and constant polymer viscosity $\mu_p$ ($D/H$=0.2, $\mu_p$=100 P). $\sigma$=0.03 mN/m.